# WiFi-Direct Simulation for INET in OMNeT++

Syphax Iskounen, Thi-Mai-Trang Nguyen and Sébastien Monnet
Sorbonne Universités, UPMC Université Paris 6, CNRS, UMR 7606, LIP6
4 Place Jussieu, 75005 Paris, France
{Syphax.Iskounen, Thi-Mai-Trang.Nguyen, Sebastien.Monnet}@lip6.fr

*Abstract*—Wi-Fi Direct is a popular wireless technology which is integrated in most of today's smartphones and tablets. This technology allows a set of devices to dynamically negotiate and select a group owner which plays the role access point. This important feature is the strength of Wi-Fi Direct and makes it more and more widely used in telecommunications networks. In this paper, we present the implementation of Wi-Fi Direct in the INET framework of OMNeT++. We have implemented the main procedures of Wi-Fi Direct such as discovery, negotiation and group formation. The implementation has been validated by two test scenarios which show the conformity of the implementation to the protocol specification.

*Keywords—Wi-Fi Direct; OMNeT++; network simulation.*

## I. Introduction

With the emergence of mobile computing and Internet of Things, new applications and services have been developed for smartphones and tablets requiring an easy way to quickly create network connection between devices. These applications usually consume more bandwidth than the traditional voice services and have been designed for data transmission between nearby devices. The concept of Device-to-Device (D2D) communication has been introduced as a means to offload the cellular macro cells and facilitate direct communications between nearby devices [1].

In the Wi-Fi technology, there were historically two communication modes in the 802.11 standard, the infrastructure mode and the adhoc mode. The infrastructure mode based on the deployment of Access Points (AP) is currently used everywhere - in a campus, offices, at home or in public hotspots - to provide Wi-Fi users with ubiquitous Internet connection. The adhoc mode allows two Wi-Fi devices to directly communicate without the necessity of having an access point. This mode is used widely as the link layer for mobile adhoc networks in conjunction with an adhoc network routing protocol in order to provide a mutihop communication. However, for several reasons such as complexity of use and lack of power saving, the adhoc mode never becomes widely deployed nor used in practical wireless networks [2]. To address the need of easily and quickly setting up D2D data communications, the Wi-Fi Alliance has developed the Wi-Fi Direct technology [3]. Today, Wi-Fi Direct is not only the de facto technology for direct communications between Wi-Fi devices but also a strong candidate for D2D communications in future 5G networks.

Wi-Fi Direct was inspired from the infrastructure mode. The devices form groups. In each group, one selected node assumes the AP functionality. The advantage of Wi-Fi Direct is the inheritance of the enhanced QoS, power saving and security mechanisms from the infrastructure mode. In addition, a large number of applications designed for connecting devices such as file sharing are available for mobile users and the APIs for Android are available for developers [4].

The research community has shown a big interest in Wi-Fi Direct. In home networking, Wi-Fi Direct enabled devices can collaborate and share streaming-based media content [5]. In vehicular networks, Android-based smartphones can use Wi-Fi Direct for fast and cheap inter-cars communications [6]. In mobile social networks, users in proximity can automatically discover each other by Wi-Fi Direct and enable a variety of interactions such as chat or games [7]. In cellular networks, the operator network can assist neighboring client devices to activate Wi-Fi Direct and communicate via the D2D links instead of cellular links in order to offload the macro cell and enhance the performance of cellular communications [1, 8].

As Wi-Fi Direct is integrated in most handheld devices, most research contributions are validated by prototyping. However, we argue that it is important for network simulators to support Wi-Fi Direct allowing researchers to evaluate the performances of the network in complex scenarios with a large number of nodes and design protocols before prototyping. In this paper, we present our implementation of Wi-Fi Direct in the INET framework version 3.2.1 of OMNeT++ version 4.6 [9], an open-source network simulator widely used in academia. The implementation consists of a new module simulating the operation of Wi-Fi Direct.

The remainder of the paper is organized as follows. In section II, we present the architecture and functionalities of Wi-Fi Direct as well as the current implementation of the IEEE 802.11 in OMNeT++. In section III, the details of Wi-Fi Direct implementation are provided. We test the implementation in section IV. Finally, we conclude the paper in section V.

## II. Background

### A. Wi-Fi Direct Architecture

Wi-Fi Direct is also called Wi-Fi Peer-to-Peer (P2P) [3]. A P2P group is a set of Wi-Fi Direct devices consisting of one P2P Group Owner (GO) and zero or more Clients as illustrated in Fig. 1.

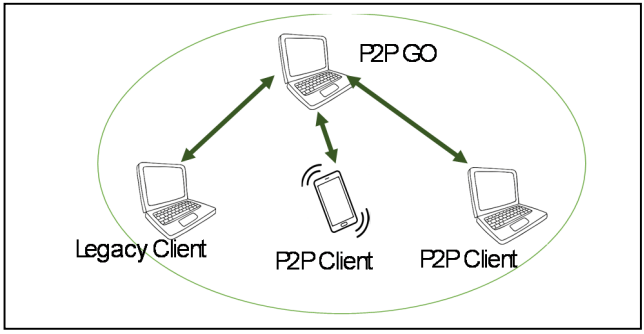

Fig. 1. Example of a Wi-Fi Direct Group.

The GO plays the role of an access point to provide clients with connectivity. It is also a communication node and can be a source or a destination of higher layer applications. All Wi-Fi Direct devices implement both roles of client and AP. These roles are dynamically assigned during the group formation. After the group formation, the P2P GO works as an AP and announces its presence by periodically sending Beacon frames. P2P devices receiving the Beacon frames can connect to the group as in a traditional Wi-Fi network. Devices that are not P2P compliant can also join the group as Legacy Clients (Fig. 1).

A P2P Group can be formed in three ways: standard, autonomous and persistent [2]. The *standard* case as shown in Fig. 2 consists of two phases, discovery and group formation.

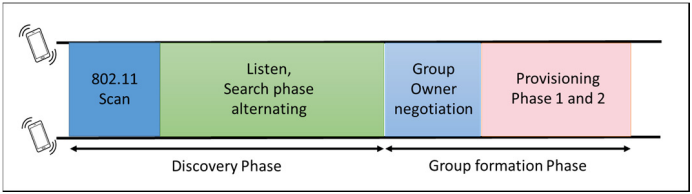

Fig. 2. Standard group formation

In the *discovery* phase, a device starts by a normal *802.11 scan* to discover existent P2P groups and traditional Wi-Fi APs. If there is no already established group, the device alternates between the *search* and *listen* states to discover another Wi-Fi Direct device which also wishes to form a group. In the search state, the device performs an active scan by sending *Probe Requests* in each channel. In the *listen* state, the device listens on a channel for a *Probe Request* to respond by a *Probe Response*. Once the two P2P devices have discovered each other by sending *Probe Request* and receiving *Probe Response* over the same channel, they can move on the group formation phase.

In the *group formation* phase, the two devices start by the *Group Owner negotiation* to know which one will become GO using a three-way handshake as illustrated in Fig. 3. It is worth noting that a Wi-Fi Direct Group may contain more than two devices but only the two first devices can discover each other and negotiate for a GO. The other devices simply detect the presence of GO and joins an existing group. In the *GO Negotiation Request* and *Response* messages, there is a *GO Intent* value declared by each participating device. The device declaring the highest value becomes the P2P GO. Once the role of GO has been agreed, the two devices switch to the *Provisioning phases 1 & 2* to establish a secure communication. In the *Provisioning phase 1*, security keys for mutual authentication are created [10]. In the *Provisioning phase 2*, the device playing the role of client reconnects to the GO using the new authentication credentials.

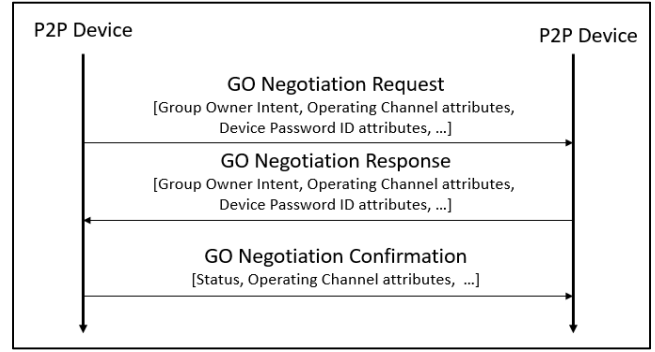

Fig. 3. Group Owner Negotiation message exchange [3]

A P2P device may create a P2P Group for itself and become immediately P2P GO by selecting a channel and sending Beacon frames. This way to form a P2P Group is called *autonomous*. Other devices can discover this Group by performing 802.11 scan and move directly on the Provisioning Phases 1 & 2 for establishing a secure communication within the group.

A group can be declared as *persistent* during first formation. A flag sent in Beacon, Probe and GO negotiation messages can perform this declaration. The devices of a persistent group store the network credentials and the role negotiated. Later on, when a device has leaved the group and wants to discover or establish a group, it can easily recognize that it has established a persistent group with the peer in the past. After the Discovery phase, the device can directly go to the Provisioning phase 2 using the previous network credentials. This way to form a P2P Group is called *persistent*.

*B. 802.11 implementation in OMNeT++*

The IEEE 802.11 infrastructure and adhoc modes have been implemented in the INET framework of OMNeT++ [11]. An IEEE 802.11 interface has four layers: agent, management, Medium Access Control (MAC) and physical layer (radio) as shown in Fig. 4.

The *physical layer* (the `Ieee80211Radio` module) models the radio propagation characteristics of the medium. The *MAC* layer (the `Ieee80211Mac` module) implements the Carrier Sense Multiple Access and Congestion Avoidance (CSMA/CA) protocol. The *management* layer implements management functions and generates management frames such as *Beacon*, *Probe*, *Authentication* and *Association* frames. This layer has several modules (e.g. `Ieee80211MgmtAdhoc`, `Ieee80211MgmtAP`, `Ieee80211MgmtSTA`) according to the role of the device (station, AP or adhoc node). The agent layer (the `Ieee80211AgentSTA` module) is currently only present for 802.11 stations and allows the user to control the behavior of a station in the network. Researchers can implement a handover strategy using this module for example.

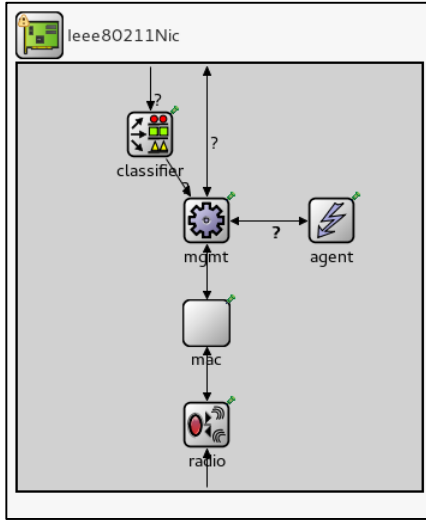

Fig. 4. IEEE 802.11 network interface module in INET Framework

### III. WI-FI DIRECT IMPLEMENTATION

In order to implement the functionalities of Wi-Fi Direct, we have created a new module at the *management* layer called `Ieee80211MgmtSTAWifiDirect` [12]. Our model is compatible with INET Framework version 2.0.0 which requires OMNeT++ version 4.2 or later because the 802.11 classes that we have modified are present from this version of the INET Framework. In fact, Wi-Fi Direct corresponds to a new communication mode of Wi-Fi networks beside the existent infrastructure and adhoc modes. An 802.11 interface plays a new role, the Wi-Fi Direct peer, which is different from the existent roles of station, AP and adhoc node. A Wi-Fi Direct peer is able to scan, discover the peer, negotiate the group owner and establish a secure communication within the group as presented in section II.

The `Ieee80211MgmtSTAWifiDirect` module performs the encapsulation and decapsulation of data frames. It exchanges management frames to perform the group formation and group joining procedures as described in Fig. 5 and Fig. 6 respectively.

Fig. 5 presents the detailed message exchange for a standard group formation. The *scan* phase corresponds to the regular IEEE 802.11 scan procedure already implemented in the `Ieee80211MgmtSTA` module. After sending Probe requests without detecting any existent GO, the Wi-Fi Direct devices alternates between the *listen* and *search* states in the Find phase until they discover each other by sending *Probe Requests* and receiving *Probe Responses* over the same channel (channel 6 in this example). Three new frames, *GO Negotiation Request/Response/Confirmation* have been implemented conforming to the protocol specification to perform the *GO negotiation* phase. For the sake of simplicity and because our internal research objective is not really related to the security aspect, we have only implemented representatively the *Provisioning phases 1 & 2* by exchanging a number of N (e.g. N = 20) authentication frames. The detailed implementation of the authentication and key exchanges in the *Provisioning* phase can be subject to a future contribution in the OMNeT++ community.

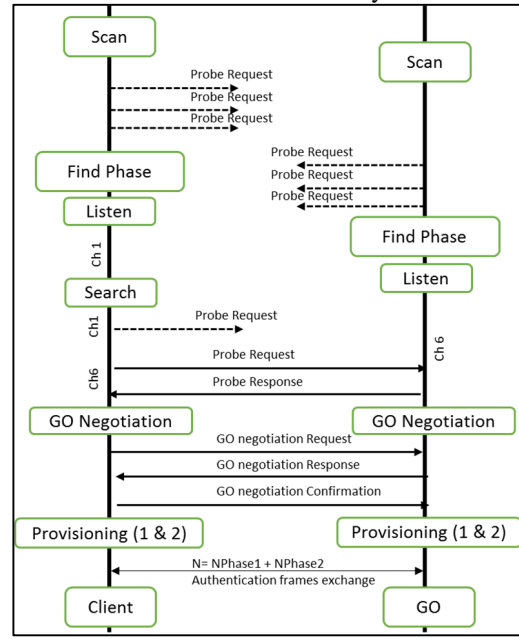

Fig. 5. Message exchange for standard group formation

Fig. 6 presents the detailed message exchange between a Wi-Fi Direct device detecting a group and the Group owner. After the scan phase (an active scan in this example), the new P2P device detects the GO of a group already formed. The P2P device asks to join the group by sending a *Provision Discovery Request*. Upon the reception of the response, it moves to the Provisioning phases 1 & 2.

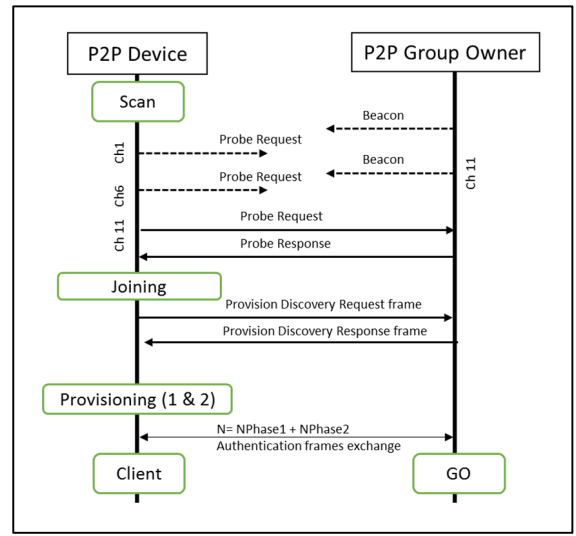

Fig. 6. Message exchange for joining an existent group

## IV. EVALUATION

In this section, we evaluate the implementation of Wi-Fi Direct by two test scenarios. In the first scenario, we create three Wi-Fi Direct hosts and let them automatically form the P2P Group following the standard group formation procedure. In the second scenario, we also create three Wi-Fi Direct host but one of them is assigned as a GO. That means this node will form a group autonomously for itself at the beginning of the simulation. As a consequence, two other hosts detect an existing GO and join the group. Sending Ping messages tests the connectivity within a group.

### A. Test of standard group formation

Fig. 7 shows the topology of the first test scenario with 3 Wi-Fi Direct hosts created.

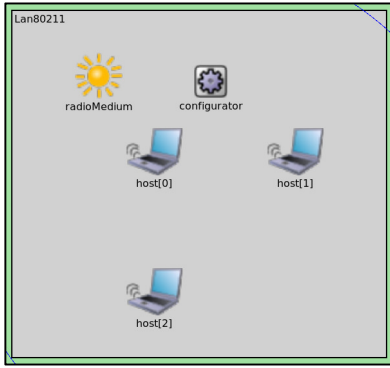

Fig. 7. Topology of Wi-Fi Direct network

In this topology we have a radio medium, an IPv4 Configurator for IP address assignment and three hosts of type `WirelessHost` with their manager module set to `Ieee80211MgmtSTAWifiDirect`. The `wirelessHost` used can be found in the INET project at `inet.node.inet.WirelessHost`.

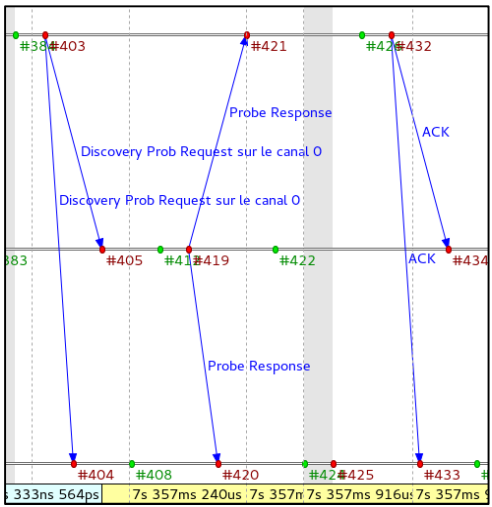

Fig. 8. Discovery phase

In the Discovery phase, each host firstly scans and doesn't detect any GO available. Then, they switch to alternating between the listen and search states to try to discover another Wi-Fi Direct host. Fig. 8 presents partially the *Discovery* phase when host[0] and host[1] discover each other. We can see that host[0] sends a *Discovery Probe Request* in channel 0 at event #403. Host[1] is in the listen state over this channel at that time and responds with a *Probe Response* message at event #419. Host[0] sends an acknowledgement (ACK) frame at event #432 and moves on the *GO negotiation* phase.

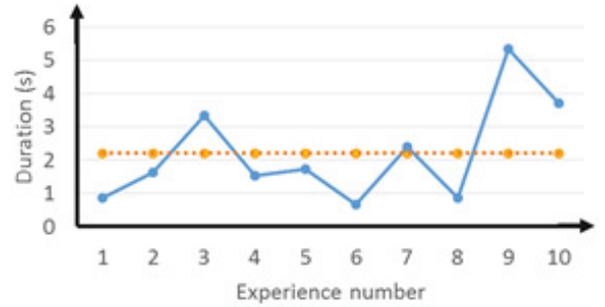

Fig. 9. Duration of the discovery phase

Fig. 9 shows the duration of the discovery phase that we have measured by our simulations. The average discovery time is about 2-3 seconds. This result corresponds the average discovery time that we have experienced with two smartphones supporting Wi-Fi Direct.

Fig. 10. The GO negotiation phase

Fig. 10 presents the GO negotiation phase with the three-way handshake between host[0] and host[1]. At the end, host[1] becomes the GO and starts sending *Beacon*s. Host[0] finishes the group formation by completing the *Provisioning phases 1 & 2* with host[1]. We have implemented these phases by the exchange of a number of *Authentication* frames which can be configured in the simulation configuration .ini file. Upon the reception of Beacon frames sent by host[1], host[2] starts the joining procedure to join the group.

Fig. 11 shows the test of connectivity within the group by sending a Ping request from host[0] to host[2] (event #2362) and another Ping request from host[1] to host[0] (event #2493). For the first Ping request, we can see that host[1] plays the role of AP relaying the Ping request from host[0] to host[2] (event #2390). The Ping reply is also transferred to host[0] via host[1], the GO of the group (event #2450). For the

second Ping request, host[0] responds by a Ping reply directly
(event #2525).

```
#2342   9.531719516371      host[1] --> host[2]     Beacon
#2362   9.548813502304      host[0] --> host[1]     ping9
#2362   9.548813502304      host[0] --> host[2]     ping9
#2375   9.549495974034      host[1] --> host[0]     ACK
#2375   9.549495974034      host[1] --> host[2]     ACK
#2390   9.549949974034      host[1] --> host[0]     ping9
#2390   9.549949974034      host[1] --> host[2]     ping9
#2407   9.550632307598      host[2] --> host[0]     ACK
#2407   9.550632307598      host[2] --> host[1]     ACK
#2422   9.551066307598      host[2] --> host[0]     ping9-reply
#2422   9.551066307598      host[2] --> host[1]     ping9-reply
#2435   9.551748641162      host[1] --> host[0]     ACK
#2435   9.551748641162      host[1] --> host[2]     ACK
#2450   9.552122641162      host[1] --> host[0]     ping9-reply
#2450   9.552122641162      host[1] --> host[2]     ping9-reply
#2465   9.552805112892      host[0] --> host[1]     ACK
#2465   9.552805112892      host[0] --> host[2]     ACK
#2493   9.592844616388      host[1] --> host[0]     ping9
#2493   9.592844616388      host[1] --> host[2]     ping9
#2510   9.593527088118      host[0] --> host[1]     ACK
#2510   9.593527088118      host[0] --> host[2]     ACK
#2525   9.593961088118      host[0] --> host[1]     ping9-reply
#2525   9.593961088118      host[0] --> host[2]     ping9-reply
#2540   9.594643559848      host[1] --> host[0]     ACK
#2540   9.594643559848      host[1] --> host[2]     ACK
#2561   9.631719516371      host[1] --> host[0]     Beacon
```

Fig. 11. Testing the connectivity for the first test scenario

*B. Test of autonomous group formation*

In the second test scenario, we use the same topology as in the first test scenario: three Wi-Fi Direct hosts, a radio medium and an IPv4 Configurator as presented in Fig. 7. The `.ini` file, presented in Fig. 12, allows us to configure host 0 as a GO so that it will perform an autonomous group formation at the beginning of the simulation with the SSID ''Wi-Fi Direct Group''.

```
# ping app host[0] pinged by Host[1]
**.numPingApps = 1
*.host[1].pingApp[0].destAddr = "host[0]"
*.host[1].pingApp[0].sendInterval = 1s
# ping app host[1] pinged by host[2]
*.host[2].pingApp[0].destAddr = "host[1]"
*.host[2].pingApp[0].sendInterval = 1s
#Configure the P2P Group
**.host[0].wlan[0].mgmt.WiFiDirectUsed=true
**.host[0].wlan[0].mgmt.WiFiDirectGO=true
**.host[0].wlan[0].mgmt.strGroup="Groupe Wifi Direct"

**.host[1].wlan[0].mgmt.WiFiDirectUsed=true
**.host[1].wlan[0].mgmt.WiFiDirectGO=false
**.host[1].wlan[0].mgmt.strGroup="Groupe Wifi Direct"

**.host[2].wlan[0].mgmt.WiFiDirectUsed=true
**.host[2].wlan[0].mgmt.WiFiDirectGO=false
**.host[2].wlan[0].mgmt.strGroup="Groupe Wifi Direct"
```

Fig. 12. Configure host 0 as GO using .ini file

We also allow the association of a host to a specific group created by an autonomous GO via the .ini file. This feature has been implemented for our internal research purpose in which we can try different group formation strategies. For example, the Wi-Fi Direct groups can be formed in a guided manner with an advanced power control technique to reduce the interference level and improve the network performances. As illustrated in Fig. 12, we configured host[1] and host[2] to be in the same group as the group created by host[0]. In practice and in other simulation scenarios, researchers can let other hosts to randomly select the group to join if there are several groups available.

```
#80    5.084426574409       host[0] --> host[1]     Beacon
#80    5.084426574409       host[0] --> host[2]     Beacon
#121   5.085398907973       host[2] --> host[0]     Provision Request
#121   5.085398907973       host[2] --> host[1]     Provision Request
#134   5.085713241537       host[0] --> host[1]     ACK
#134   5.085713241537       host[0] --> host[2]     ACK
#149   5.086107713267       host[1] --> host[0]     Provision Request
#149   5.086107713267       host[1] --> host[2]     Provision Request
#162   5.086422184997       host[0] --> host[1]     ACK
#162   5.086422184997       host[0] --> host[2]     ACK
#177   5.086796184997       host[0] --> host[1]     Provision discovery Response
#177   5.086796184997       host[0] --> host[2]     Provision discovery Response
#190   5.087110518561       host[2] --> host[0]     ACK
#190   5.087110518561       host[2] --> host[1]     ACK
#205   5.087504852125       host[0] --> host[1]     Provision discovery Response
#205   5.087504852125       host[0] --> host[2]     Provision discovery Response
#218   5.087819323855       host[0] --> host[1]     ACK
#218   5.087819323855       host[1] --> host[2]     ACK
```

Fig. 13. Autonomous group formation and joining procedure

As presented in Fig. 13, host[0] starts to send *Beacon* frames when we launch the simulation to announce its presence. The other hosts, at the reception of the *Beacon* frames sent by host[0], start the joining procedure by sending the *Provision Discovery Request* frame. The P2P GO (host[0]) replies with the *Provision Discovery Response* frame. After the joining procedure, the *Provisioning phases 1 & 2* happen (that we do not show here) as usual to accomplish the group formation.

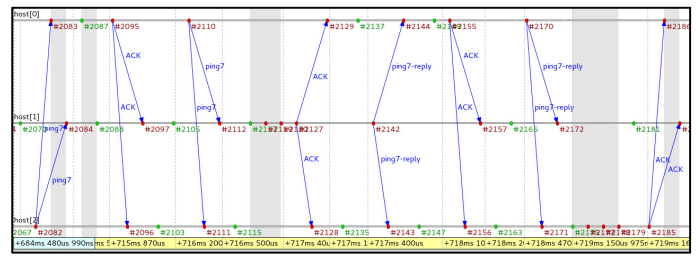

Fig. 14. Connectivity test for scenario 2

Fig. 14 shows the connectivity test for the group. Host[2] sends a Ping request to host[1] via host[0], the GO. Similarly, the Ping reply sent from host[1] to host[2] is also relayed by the GO.

V. CONCLUSION

In this work, we have implemented Wi-Fi Direct in the INET Framework of OMNeT++ [12]. The functionalities of Wi-Fi Direct are implemented as a management module in the same way as other IEEE 802.11 management modules already existing in the INET Framework in order to facilitate its utilization, its customization and its integration in OMNeT++. This implementation can be used for research on networking using Wi-Fi Direct such as protocol design and performance evaluations. In future works, we will integrate our modules into the INET Framework and use them for the research on Wi-Fi Direct-based device-to-device communications in dense wireless networks to offload the Wi-Fi network infrastructure.

REFERENCES

[1] A. Asadi and V. Mancuso, "WiFi Direct and LTE D2D in action", IFIP Wireless Days (WD), Valencia, Spain, November 2013.
[2] D. Camps-Mur, A. Garcia-Saavedra and P. Serrano, « Device-to-Device Communications with WiFi Direct:

Overview and Experimentation », IEEE Wireless Communications, Vol. 20, No. 3, June 2013.
[3] Wi-Fi Alliance, "Wi-Fi Peer-to-Peer (P2P) Technical Specification v1.5", August 2014.
[4] Wi-Fi Alliance, URL: http://www.wi-fi.org
[5] H. Yoon and J. Kim, "Collaborative Streaming-based Media Content Sharing in WiFi-enabled Home Networks", IEEE Transactions on Consumer Electronics, Vol. 56, No. 4, November 2010.
[6] A. Djajadi and R.J. Putra, "Inter-cars Safety Communication System Based on Android Smartphone", IEEE Conference on Open Systems (ICOS), Subang, Malaysia, October 2014.
[7] J. Zuo, Y. Wang, Q. Jin and J. Ma, "HYChat: A Hybrid Interactive Chat System for Mobile Social Networking in Proximity", IEEE International Conference on Smart City (SmartCity), Chengdu, China, December 2015.
[8] A. Pyattaev, K. Johnsson, A. Surak, R. Florea, S. Andreev and Y. Koucheryavy, "Network-assisted D2D Communications: Implementing a Technology Prototype for Cellular Traffic Offloading", IEEE Conference on Wireless Communications and Networking (WCNC), Istanbul, Turkey, April 2014.
[9] INET Framework, URL: https://inet.omnetpp.org/
[10] Wi-Fi Alliance, « Wi-Fi Simple Configuration Technical Specification, Version 2.0.5 », August 2014.
[11] OMNeT++, « INET Framework Manual – Chapter 9: The 802.11 Model", January 2016.
[12] Wi-Fi Direct Implementation Source Code. URL: http://www-phare.lip6.fr/~trnguyen/research/wifidirect